\documentclass[]{spie}
\usepackage[]{graphicx}

\title{Forward Diffracted Parametric X Radiation
From a Thick Tungsten Single Crystal at 855 MeV Electron Energy}

\author{H. Backe\supit{a}, W. Lauth\supit{a}, A.F. Scharafutdinov\supit{a},
P. Kunz\supit{a},
\\A.S. Gogolev\supit{b}, A.P.
Potylitsyn\supit{b} \skiplinehalf \supit{a}Institut f\"{u}r
Kernphysik, Johannes
Gutenberg-Universit\"{a}t Mainz, \\D-55099 Mainz, Germany\\
\supit{b}Institute for Nuclear Physics, Tomsk Polytechnic
University, \\634004 Tomsk, Russian Federation}

\authorinfo{Send correspondence to H. Backe\\
E-mail: Backe@kph.uni-mainz.de, Telephone: ++49-6131-39-25563}

\begin{document}
\maketitle

\begin{abstract}
Features of forward diffracted Parametric X-Radiation (PXR) were
investigated at experiments with the 855 MeV electron beam of the
Mainz Microtron MAMI employing a 410 $\mu$m thick tungsten single
crystal. Virtual photons from the electron field are diffracted by
the (10$\bar{1}$) plane at a Bragg angle of 3.977$^\circ$. Forward
emitted radiation was analyzed at an energy of 40 keV with the
(111) lattice planes of a flat silicon single crystal in Bragg
geometry. Clear peak structures were observed in an angular scan
of the tungsten single crystal. The results were analyzed with a
model which describes forward diffracted PXR under real
experimental conditions. The experiments show that forward
diffracted PXR may be employed to diagnose bending radii of
lattice planes in large area single crystals.
\end{abstract}


\keywords{Parametric X radiation}

\section{INTRODUCTION} \label{sect:intro}
If a charged particle passes an interface between two amorphous
media with different dielectric susceptibilities broad band
electromagnetic radiation is produced. The frequency spectrum of
this so-called transition radiation (TR) has been calculated by
Ginsburg and Frank \cite{GinF45} and Garibian \cite{Gar58,Gar61}.
In the ultra relativistic case the radiation distribution
resembles a Lorentz transformed dipole pattern with maximum
intensity at an angle 1/$\gamma$ with $\gamma=1/\sqrt{1-\beta^2}$
the relativistic factor, $\beta=v/c$, $v$ the electron velocity,
and $c$ the speed of light. The energy spectrum extends up to a
gradual cut-off energy $\gamma\hbar\omega_p$ with $\omega_p$ the
plasma frequency. For an electron energy of 855 MeV, as available
at the Mainz Microtron MAMI, the characteristic opening angle at
$\gamma$~=~1673 amounts to 0.6 mrad. The cut off energy for
tungsten with $\hbar \omega_p$ = 80.4 eV is as high as 135 keV.

When a relativistic electron crosses the boundary between vacuum
and a single crystal the TR propagates into the crystal in forward
direction. A diffracted transition radiation (DTR) reflex with a
wave vector $\textbf{k}_r$ is created if the wave vector
$\textbf{k}_v$ associated with the virtual photon field of the
electron fulfils approximately the well known Bragg law in vector
form $\textbf{k}_v+\textbf{H}\simeq\textbf{k}_r$, with
$\textbf{H}$ a reciprocal lattice vector of a specific crystal
plane, see e.g. \cite{Cat89}. Inside the crystal the TR field will
be extinguished by Bragg scattering, photo absorption or Compton
scattering. After a sufficiently long distance from the entrance
interface its amplitude may become negligibly small. Finally, deep
inside the crystal, only the virtual photon field associated with
the electron remains. But also under these circumstances
monochromatic X rays are emitted close to the Bragg direction.
This kind of radiation was predicted in the framework of a
kinematical theory by Ter-Mikaelian \cite{Ter72} in which only one
diffracted wave is required. Baryshevsky and Feranchuk
\cite{Bar71,BarF71}, and Garibian and Yang \cite{GarY71,GarY72}
explained this so-called parametric X-ray radiation (PXR) with
dynamical theories in which an additional forward diffracted wave
occurs. According to theoretical predictions PXR is
quasi-monochromatic and features at a fixed observation angle
narrow energy band characteristics. It was suggested that PXR is
just a kind of \v{C}erenkov radiation, see Ref. \cite{GarY86} and
citations therein. Later on, kinematical theories were developed
by Nitta \cite{Nit91,Nit92}, Achim Richters group \cite{FreG95},
and dynamical ones by Caticha \cite{Cat89,Cat92} and Artru et al.
\cite{ArtR01}. Experimentally PXR was studied in a number of
papers, see e.g. for an overview up to the year 1997 Ref.
\cite{BreH97} and references cited therein, as well as Ref.
\cite{FreG97, BreL97, MorS97, FreG00}.

It is very difficult to decide experimentally whether PXR emission
is a kinematical or dynamical process. The reason has been
discussed by Nitta \cite{Nit00}. He showed that the first-order
approximation of the dynamical calculation gives the kinematical
expression. Extremely accurate absolute intensity measurements
would be required to detect a difference. Baryshevsky \cite{Bar97}
proposed to search for the predicted forward diffracted wave which
is associated to PXR and emitted close to the direction the
electron travels. Similar proposals have also been communicated by
Nasonov \cite{KubN03,NasN03}. A number of unsuccessful experiments
were performed for the search of such a forward diffracted
parametric X radiation (FDPXR) \cite{YuaA85,KalN01,BacB94,KerT98}.
First observations of interference phenomena in the TR emission
spectrum close to a Bragg energy from thin silicon single crystal
targets were reported by the Mainz group \cite{BacA03}. Later on,
the forward diffracted PXR peaks were observed by the
Tomsk-Belgorod \cite{AleB04} and the Mainz \cite{BacR05} groups.

In this contribution we describe results obtained with a 410
$\mu$m thick tungsten single crystal at the Mainz Microtron MAMI.
This crystal, cut with the $[111]$ direction perpendicular to the
surface, was also used for experiments of Ref. \cite{AleB04}. In
the experiments described in this contribution special emphasis
was put on a more quantitative understanding of the forward
diffracted PXR peak for which also the influence of the analyzing
crystal spectrometer device on the signal generation had to be
studied carefully.

\section{BASICS}\label{sect:basics}
\subsection{Parametric X Radiation as a Dynamical Process}
Let us consider the radiation fields of a highly relativistic
electron in a semi-infinite crystal. The electron is assumed to
move on a straight trajectory with constant relativistic factor
$\gamma$. If the electron crosses the boundary between vacuum and
crystal transition radiation (TR) and transition diffracted
radiation (TDR) is created, see e.g. \cite{Cat89}. The TR field
propagates in the crystal in forward direction, the TDR field is
assumed to be associated with a Laue reflection characterized by a
reciprocal lattice vector $\textbf{H}$. Both fields will be
attenuated in the crystal by photo absorption or Compton
scattering and after a sufficient long distance from the entrance
interface their amplitudes may become negligibly small. Finally,
only the virtual photon field ${A}_{c}$ associated to the electron
remains. We are interested in the interaction of this field with
the crystal which can best be discussed in the framework of
dispersion surfaces. Following the work of Caticha \cite{Cat89} in
Fig. 1 dispersion surfaces of the electron field (DSEF) and of
X-rays in the crystal (DSXC) are shown. In this picture a virtual
photon is characterized by a wave vector $\textbf{k}_v$, starting
at a point V of the DSEF and pointing to the origin O, and the
field amplitude ${A}_{c0}$. A particular situation happens if the
point V approaches the intersection C between the DSEF and the
DSXC$_H$ with its origin at the point H. In the surrounding of
this intersection point C the virtual photon field of the electron
is diffracted. The momentum transfer $\hbar \textbf{H}$ transforms
$ \textbf{k}_{v}$ into $\textbf{k}_{H}=\textbf{k}_{v}+ \textbf{H}$
and a real photon in the crystal with a field amplitude ${A}_{cH}$
is created. This is the well known PXR field. The corresponding
forward diffracted field ${A}_{c0}$, however, remains a virtual
one. A momentum transfer $ \Delta p_{c}=(\hbar K/2)(1/\gamma^2+
\vartheta_{v}^2-\chi_{0})$ would be required to shift it on the
DSXC$_0$ which has its origin at point O. In this equation $
K={|\textbf{k}_{r}|}=\omega/c$ is the vacuum wave vector of the
radiation field, $\overrightarrow{\vartheta_{v}}$ the angle
between $ \textbf{v}$ and $ \textbf{k}_{v}$, and
$\chi_{0}=\chi_{0}^{\prime} + i\chi_{0}^{\prime\prime}$ the mean
dielectric susceptibility of the crystal.
\begin{figure}
   \begin{center}
   \begin{tabular}{c}
   \includegraphics[height=10cm]{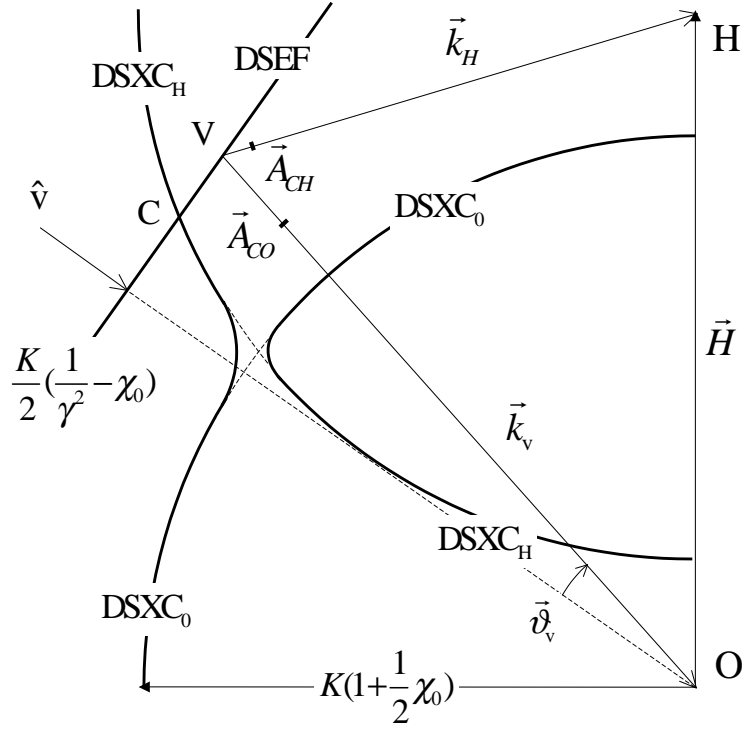}
   \end{tabular}
   \end{center}
   \caption[]
{Dispersion surfaces. Shown are the dispersion surface of the
electron field (DSEF) and of X-rays in the crystal (DSXC). The
quantity $\textbf{k}_{v}$ is the wave vector of the virtual field
amplitude ${A}_{c0}$ of the electron,
$\textbf{k}_{H}=\textbf{k}_{v}+ \textbf{H}$ the wave vector of the
real PXR field ${A}_{cH}$\label{figDispSurf}}
\end{figure}
The fact that only one real field amplitude ${A}_{cH}$ exists in
the crystal at the intersection C between DSEF and DSXC$_H$
resembles a kinematical picture of PXR generation. But in reality
the virtual field of the electron is diffracted into two field
amplitudes ${A}_{c0}$ and ${A}_{cH}$ from which the former remains
a virtual field. It must be concluded that in the framework of
this picture PXR production is a dynamical process.

It remains to be discussed how the existence of the forward
diffracted virtual field ${A}_{c0}$ at the intersection C can be
established experimentally. Of course, as a virtual field it can
not be observed directly. However, let us assume that the electron
exits the crystal at a second boundary again into vacuum. The
continuity of the vector potential at this surface requires an
additional radiation field in vacuum ${A}_{r}={A}_{c0}- {A}_{v}$
with ${A}_{v}$ the virtual field amplitude of the moving electron
in vacuum. At a point V far off the intersection point C the
emitted radiation is just again the well known TR. However, close
to the intersection C the additional forward diffracted field
amplitude will contribute in addition. The free X-ray field
${A}_{r}$ can be observed with an angular and energy resolving
radiation detector system. It is expected that the amplitude
${A}_{c0}$ and therefore also the radiation field ${A}_{r}$
exhibits a resonance behavior at well defined correlated photon
emission angles $\overrightarrow{\vartheta}$ and photon energies
$\hbar\omega$ which we call in the following forward diffracted
parametric X radiation (FDPXR). As will be discussed elsewhere in
more detail, FDPXR can either be observed as an "amplitude
contrast" from a thick (semi-infinite) crystal or as a "phase
contrast" from a thin crystal slab. In the former only the
amplitude from the exit interface remains, while in the latter the
radiation amplitude created at the entrance interface survives and
interferes coherently with the amplitude created at the exit
interface. Both experiments have been performed at the Mainz
Microtron MAMI \cite{BacR05}.

\subsection{Theoretical Background}\label{subsect:theory}
Following the formalism of Caticha \cite{Cat89}, the amplitude
$A_{rP}^0$ of the radiation in forward direction and $A_{rP}^H$
for PXR, emitted from an electron exiting a semi-infinite crystal,
are given by the Equations
\begin{eqnarray}
\frac{c}{e}K_{0}^2A_{rP}^0=
-\frac{\hat{\textbf{v}}\cdot\hat{\textbf{e}}_P}{|\hat{\textbf{v}}\cdot\hat{\textbf{n}}|}
\frac{\omega}{4v}\bigg(Z_{v}-Z_{a}+Z_{a}
\frac{(\varepsilon_{1}-i\gamma_{1}/2)P^2}
{\varepsilon+\varepsilon_{0}-i\gamma_{0}/2+(\varepsilon_{1}-i\gamma_{1}/2)P^2}\bigg)\label{arp0}
\end{eqnarray}
and
\begin{eqnarray}
\frac{c}{e} K_{0}^2 A_{rP}^H= -\frac{\hat{\textbf{v}} \cdot
\hat{\textbf{e}}_P}{|\hat{\textbf{v}} \cdot \hat{\textbf{n}}|}
~\frac{\omega}{4v}\bigg(Z_{a}
\frac{P\chi_{H}/(4\sin^2(\Theta_{0}))}{\varepsilon+\varepsilon_{0}-i\gamma_{0}/2+
(\varepsilon_{1}-i\gamma_{1}/2)P^2}\bigg),\label{arpH}
\end{eqnarray}
respectively. Here
\begin{eqnarray}
Z_{v}=\frac{4v}{\omega}~\frac{1}{1/\gamma^2+(\vec{\vartheta}-\vec{\varphi})^2}\label{Zv}
\end{eqnarray}
and
\begin{eqnarray}
Z_{a}=\frac{4v}{\omega}~\frac{1}{1/\gamma^2+(\vec{\vartheta}-\vec{\varphi})^2-\chi_{0}}\label{Zas}
\end{eqnarray}
are the vacuum formation length and the mean formation length in
the crystal, respectively. Notice that the latter is just the
formation length for amorphous matter. The quantity
$\varepsilon=(\hbar\omega-\hbar\omega_0)/\hbar\omega_0$ is the
relative energy deviation from the reference energy
$\hbar\omega_0=\hbar c H_0/(2 \sin\Theta_0)$ with $H_0 = 2 \pi
\sqrt{h^2+k^2+l^2}/a_0$, $h, k, l$ the Miller indices, and $a_0$
the lattice constant. The quantities $\varepsilon_0$,
$\varepsilon_1$, $\gamma_0$, and $\gamma_1$ are functions of
various geometrical variables as the observation angle
$\vec{\vartheta}$, the deviation $\vec{\varphi}$ of an individual
electron from the nominal direction which coincides with the $z$
axis, the rotation angle $\psi$ around the vertical $y$ axis which
describes a small deviation of the reciprocal lattice vector
$\textbf{H}$ from the nominal orientation $\textbf{H}_0$ in the
$(x,z)$ plane, and the Fourier components $\chi_0$ and $\chi_H$ of
the dielectric susceptibility of the crystal, see Ref.
\cite{LucS91}. For our experimental conditions these quantities
are sufficiently well approximated by
\begin{eqnarray}
\varepsilon_{0}&=&\frac{\vartheta_{x}-\psi}{\tan\Theta_{0}}
-\frac{1/\gamma^2+(\vec{\vartheta}-\vec{\varphi})^2-\chi^{\prime}_{0}}{4\sin^2\Theta_{0}},
\label{e0}\\
\varepsilon_{1}&=&\frac{\chi^{\prime 2}_{H}}
{[1/\gamma^2+(\vec{\vartheta}-\vec{\varphi})^2-\chi^{\prime}_{0}]^24\sin^2\Theta_{0}},\label{e1}\\
\frac{\gamma_{0}}{2}&=&\frac{-\chi^{\prime\prime}_{0}}{4\sin^2\Theta_{0}},\label{gamma0}\\
\frac{\gamma_{1}}{2} &=& \frac{ -\chi^{\prime 2 }_{H}
\chi^{\prime\prime}_{0}-2\chi^{\prime}_{H}\chi^{\prime\prime}_{H}
[1/\gamma^2+(\vec{\vartheta}-\vec{\varphi})^2-\chi^{\prime}_{0}]}{
[1/\gamma^2+(\vec{\vartheta}-\vec{\varphi})^2-
\chi^{\prime}_{0}]^2\,4\sin^2\Theta_{0}}.\label{gamma1}
\end{eqnarray}
Moreover, the factor $P$ is 1 or $\cos(2\Theta_0)$ for $\sigma$ or
$\pi$ polarization, respectively, $\hat{\textbf{e}}_P$ the desired
polarization state, $K_0~=~\omega_0/c$, and e the charge of the
electron. Finally, the polarization factor $\hat{\textbf{v}}
\hat{\textbf{e}}_{P}$ must be calculated. We distinguish between
$\pi$ polarization for which the polarization vector
$\hat{\textbf{e}}_{\pi}$ lies in the plane spanned by the vector
$\textbf{k}_{v}$ and $\textbf{H}$, and $\sigma$ polarization for
which $\hat{\textbf{e}}_{\sigma}$ is perpendicular to this plane.
Notice, that in both cases the polarization vector
$\hat{\textbf{e}}_P$ is perpendicular to the unit wave vector
$\textbf{k}_v$ of the virtual photon. The result is
\begin{eqnarray}
 \hat{\textbf{v}} \hat{\textbf{e}}_{P} = \left\{\begin{array}{ll}
         \frac{(\vec{\vartheta}-\vec{\varphi})\cdot\hat{\bf{H}_0}}{
\cos(\Theta_{0})} & \mbox{for $\pi$ polarization with $P=\cos2\Theta_{0}$};\\
 & \\
        -\frac{(\vec{\vartheta}-\vec{\varphi})\cdot(\hat{\textbf{e}}_{z}\times
\hat{\bf{H}_0})}{\cos(\Theta_{0})} & \mbox{for $\sigma$
polarization with P = 1}.\end{array} \right. \label{vep}
\end{eqnarray}

It is important to realize that the counterpart of the
"diffracted" PXR amplitude, Eq. (\ref{arpH}), can be found in Eq.
(\ref{arp0}) in the "primary" FDPXR amplitude
\begin{eqnarray}
\frac{c}{e}K_{0}^2A_{rP}^{0H}=
-\frac{\hat{\textbf{v}}\cdot\hat{\textbf{e}}_P}{|\hat{\textbf{v}}\cdot\hat{\textbf{n}}|}
\frac{\omega}{4v}\bigg(Z_{a}
\frac{(\varepsilon_{1}-i\gamma_{1}/2)P^2}
{\varepsilon+\varepsilon_{0}-i\gamma_{0}/2+(\varepsilon_{1}-i\gamma_{1}/2)P^2}\bigg).
\label{A0H}
\end{eqnarray}
PXR and FDPXR amplitudes are intimately connected with each other.
Both amplitudes have the same structure, in particular their poles
are identical, and both amplitudes disappear for amorphous matter,
i.e. for $\chi_H$ = 0. In this case the parameters $\varepsilon_1$
and $\gamma_1$ are zero, and the forward amplitude, Eq.
(\ref{arp0}), reduces to the expression for transition radiation
from a single interface of amorphous matter. The same is true far
off the resonance where the PXR and FDPXR amplitudes are
negligibly small.
\begin{figure}
   \begin{center}
   \begin{tabular}{c}
   \includegraphics[width=13.0 cm]{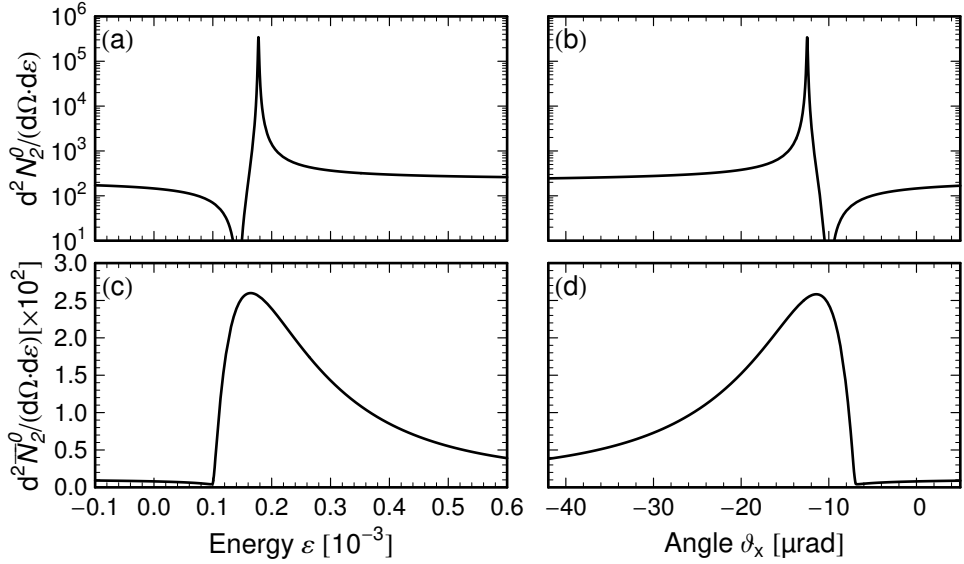}
   \end{tabular}
   \end{center}
   \caption[]
{Calculated forward diffracted intensity distributions for the
target in positive orientation, see Fig.
\ref{figmodExpPlanePlusMinus} (a) for explanation. Panel (a) shows
calculations with Eq. (\ref{d2N02phiout}) in a logarithmic scale.
A resonance is seen as function of the relative photon energy
$\varepsilon=\Delta\hbar\omega/\hbar\omega_{0}$ for observation in
beam direction $\vartheta_x$ = $\vartheta_y$ = 0 mrad, rotation
angle $\psi$ = 0 mrad. Panel (b) shows the resonance as function
of the observation angle $\vartheta_x$ ($\vartheta_y$ = 0 mrad)
for $\varepsilon = 0$, electron beam direction $\varphi_x$ = 1.0
mrad, $\varphi_y$ = 0 mrad, and rotation angle $\psi$ = 0 mrad.
The tungsten radiator single crystal is assumed to be cut with the
(111) plane parallel to the surface. Reciprocal lattice vector
$\hat{\textbf{H}_0}$= (0.997592, 0, -0.0693592), and surface
normal $\hat{\textbf{n}_0}$= (0.0693592, 0, 0.997592) were
assumed. The Bragg angle for the $(10\bar{1})$ reflection is
$\Theta_{0}=3.977 ^{\circ}$ resulting with a lattice constant
$a_{0}=3.16 \AA$ in a photon energy $\hbar\omega_{0}=40$~keV.
Fourier components of the dielectric susceptibility are
$\chi^{\prime}_{0}=-0.399 \times 10^{-5}$,
$\chi^{\prime\prime}_{0}=-0.923\times 10^{-7}$,
$\chi^{\prime}_{H}=-0.325\times 10^{-5}$,
$\chi^{\prime\prime}_{H}=-0.916\times 10^{-7}$ \cite{LucS91}.
Panels (c) and (d) show corresponding calculations of panels (a)
and (b) by Eq. (\ref{d2N02phioutmeanP}) and
(\ref{d2N02phioutmeanS}) in a linear scale for a Gaussian-like
scattering distribution of the electrons at an rms value $\sigma$
= 5.32 mrad. Shown is the sum over both polarization states.}
\label{figResonance}
   \end{figure}

If the polarization state is not observed the final result for the
total number of photons $d^2N^0$ emitted per relative energy band
width $d\varepsilon = d\hbar \omega/\hbar \omega_{0}$ into the
solid angle $d\Omega$ is the incoherent sum of the $\pi$ and
$\sigma$ polarization contributions and reads
\begin{eqnarray}
\frac{d^2N^0_{2}}{d\Omega
d\varepsilon}(\psi,\vec{\varphi},\vec{\vartheta},\varepsilon)=
\sum_{P}\frac{d^2N^0_{2P}}{d\Omega
d\varepsilon}(\psi,\vec{\varphi},\vec{\vartheta},\varepsilon)=
\frac{\alpha}{\pi^2}\cdot
|\hat{\bf{n}}\cdot\hat{\bf{k}_{r}}|^2|\,\frac{c}{e}K_{0}^2\,|^2\sum_P
|\,A^0_{rP}(\psi,\vec{\varphi},\vec{\vartheta},\varepsilon)\,|^2.\label{d2N02phiout}
\end{eqnarray}
The intensity distribution of the forward emitted radiation for a
single interface is depicted in Fig. \ref{figResonance}. Panel (a)
resembles the cross-section of resonance s-wave neutron scattering
off heavy nuclei. Similar as in the latter, the smooth part of the
formation length $Z_v-Z_a$ and the additional FDPXR resonance term
interfere destructively at the low energy side of the resonance
and constructively at the high energy side. The resulting
structure has a very narrow width, in the order of 8 meV only.
However, any angular distribution of the electrons at the exit of
the crystal in the angle $\vec{\varphi}$ may deteriorate the line
width. Assuming a Gaussian with an rms value $\sigma$ = 5.32 mrad,
which is expected from multiple scattering of the electrons in a
tungsten crystal with a thickness of 410 $\mu$m \cite{LynD91}, the
intensity distributions of each polarization state are the
integrals
\begin{eqnarray}
\frac{d^2\overline{N}^0_{2\pi}}{d\Omega d\varepsilon}
\big(\psi,\vec{\vartheta},\varepsilon\big)&\cong&\frac{1}{2
\sigma^2}\int\limits_{0}^{\infty}
{\varphi~e^{-\varphi^2/(2\sigma^2)}\frac{d^2N^0_{2\pi}}{d\Omega
d\varepsilon}\big(\psi,\varphi_{x}=\varphi,\varphi_{y}=0,\vec{\vartheta},\varepsilon\big)~d\varphi},
\label{d2N02phioutmeanP}\\
\frac{d^2\overline{N}^0_{2\sigma}}{d\Omega d\varepsilon}
\big(\psi,\vec{\vartheta},\varepsilon\big)&\cong&\frac{1}{2
\sigma^2}\int\limits_{0}^{\infty}
{\varphi~e^{-\varphi^2/(2\sigma^2)}\frac{d^2N^0_{2\sigma}}{d\Omega
d\varepsilon}\big(\psi,\varphi_{x}=0,\varphi_{y}=\varphi,\vec{\vartheta},\varepsilon\big)~d\varphi}.
\label{d2N02phioutmeanS}
\end{eqnarray}
In this approximation the double integral over $\varphi_{x}$ and
$\varphi_{y}$ has been replaced by a single integral. The
approximation is supposed to be sufficiently accurate as long as
the observation angle $\vec{\vartheta}$ is very small in
comparison to $\sigma$. For $\overrightarrow{\vartheta} = 0$ the
angles $\varphi_{x}$ and $\varphi_{y}$ enter in Eq.
(\ref{d2N02phiout}) symmetrically in second order as
$\varphi_{x}^2+\varphi_{y}^2 = \varphi^2$ with the exception of a
pre-factor $\varphi_{x}^2$ and $\varphi_{y}^2$ for $\pi$ and
$\sigma$ polarization, respectively, originating from Eq.
(\ref{vep}). With the substitution $\varphi_{x} = \varphi
\cos\alpha$ the integration can be carried out in polar
coordinates. The integral over $\alpha$ results in a common factor
$\pi$ which cancels in Eq. (\ref{d2N02phioutmeanP}) and
(\ref{d2N02phioutmeanS}) with a factor $\pi$ in the de-nominator
originating from the normalization of the two-dimensional
Gaussian. The result is shown in Fig. \ref{figResonance} (c) in a
linear scale. The line is still as narrow as about 8 eV or 2$\cdot
10^{-4}$ with respect to the photon energy but has rather long
tails. The reason of this quite unexpected result can again be
found in the already mentioned fact that the quantities
$(\vartheta_{x}-\varphi_{x})$ and $(\vartheta_{y}-\varphi_{y})$
enter in Eq. (\ref{d2N02phiout}) in second order, see Eqns.
(\ref{e0}), (\ref{e1}), (\ref{gamma0}), (\ref{gamma1}), and
\ref{vep}. Panels (b) and (d) show corresponding results of the
intensity distribution at a fixed photon energy. It is noteworthy
that even with the scattering distribution with an angular rms
value of 5.32 mrad the angular width $\vartheta_x$ of the
resonance amounts to only 14 $\mu$rad.

The detection of such narrow structures is difficult. While the
regular PXR peak is background free and can be observed with
detectors of moderate energy resolution, the FDPXR structure is
always located on top of the smooth transition radiation
background. This TR background can clearly be seen in Fig.
\ref{figResonance} (a) and (b). To find the narrow structures
experimentally this background must be suppressed which requires
for the detector a comparably very good energy resolution. As will
be described in the next subsection, a silicon single crystal
monochromator is well suited for this purpose.
\begin{figure}
   \begin{center}
   \begin{tabular}{c}
   \includegraphics[width=10.0 cm,clip]{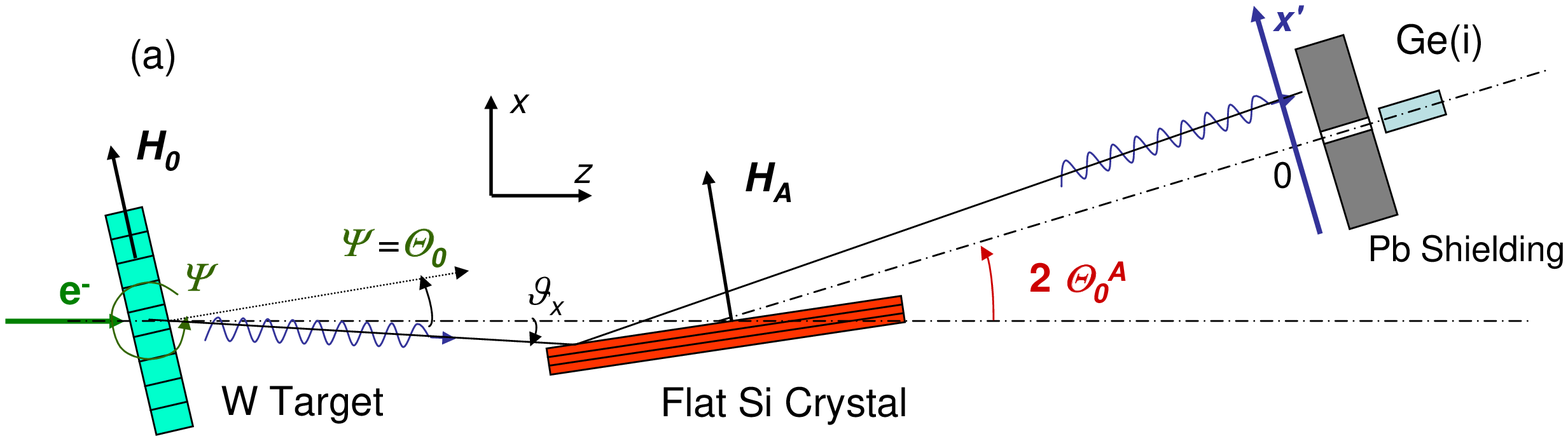}\\
   \includegraphics[width=14.5 cm,clip]{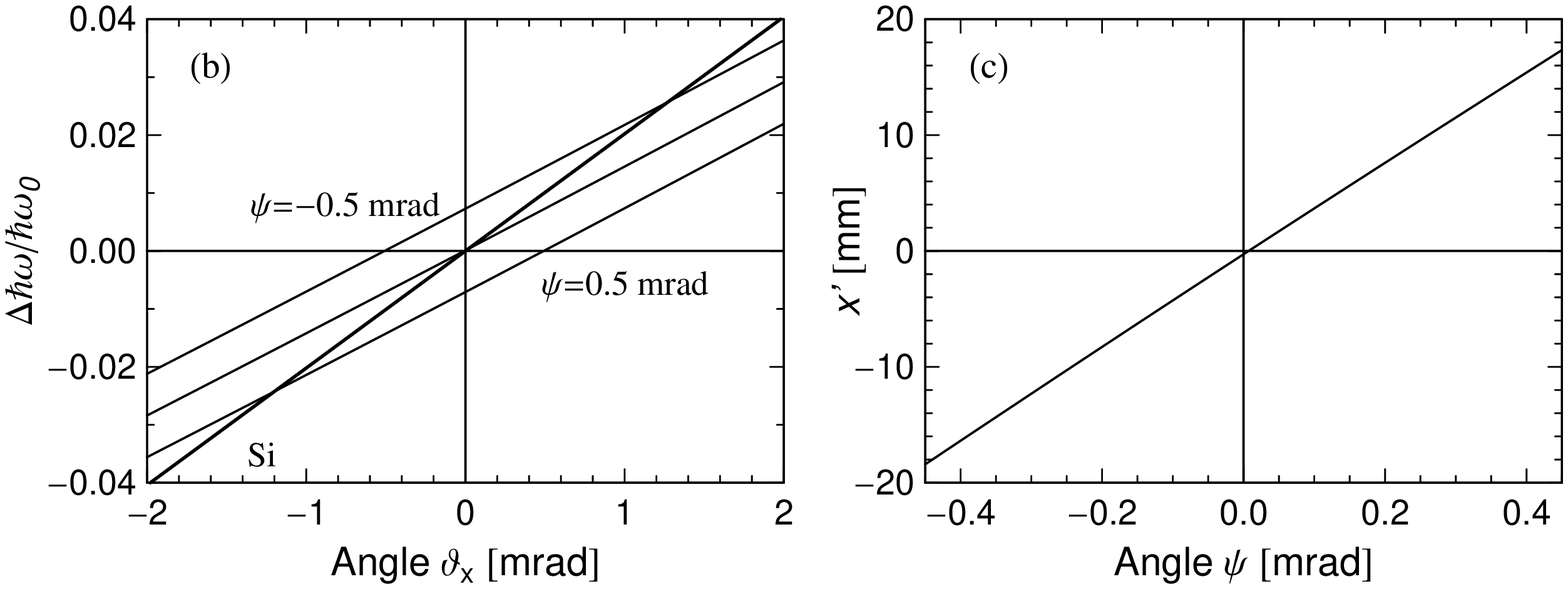}\\
   \\
   \\
   \includegraphics[width=10.0 cm,clip]{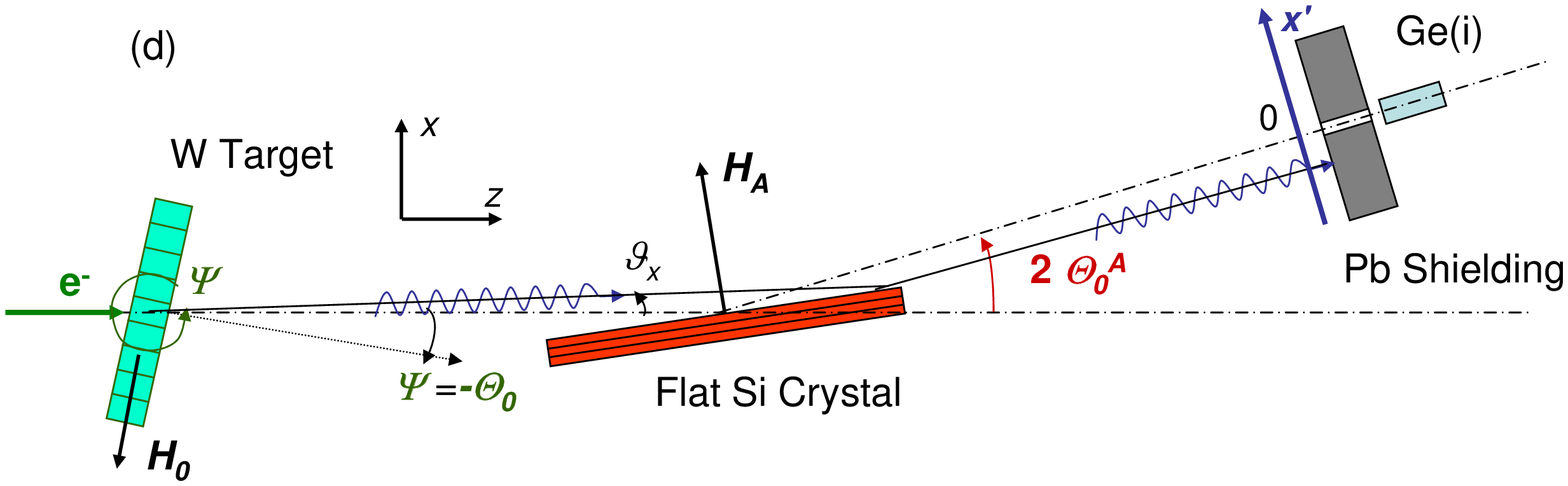}\\
   \includegraphics[width=14.5 cm]{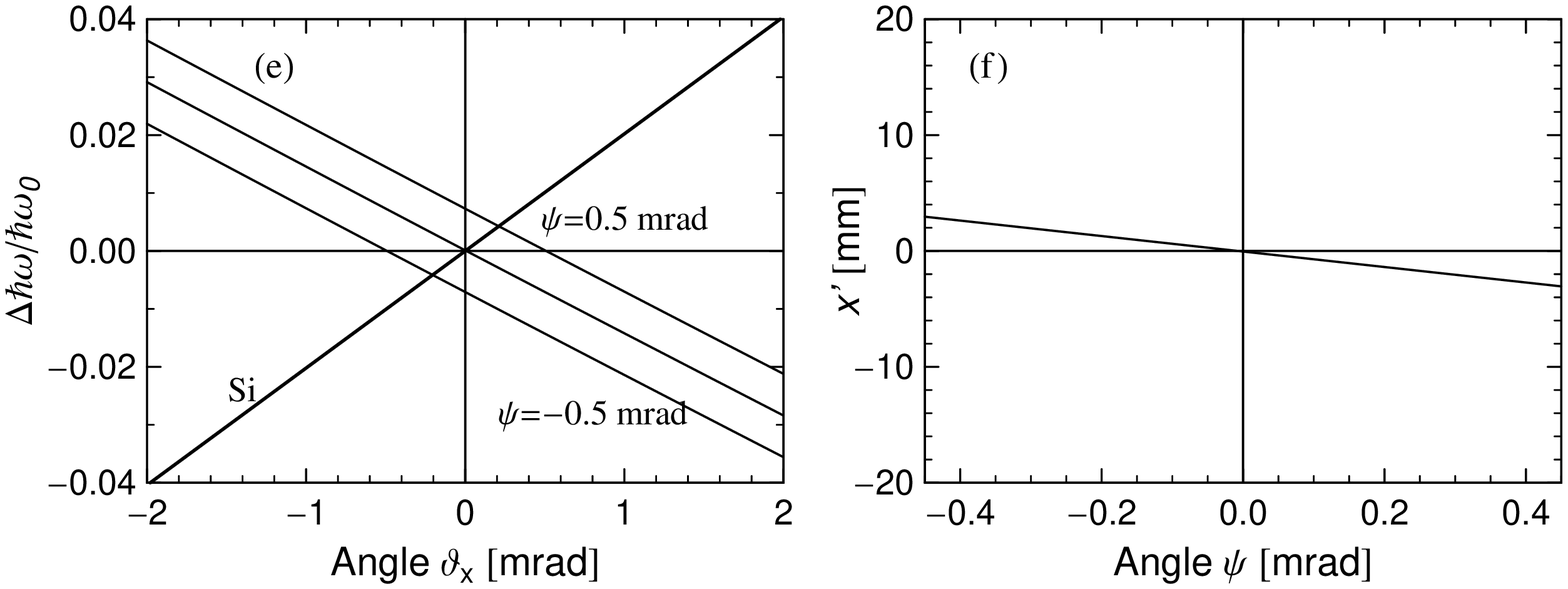}
   \end{tabular}
   \end{center}
   \caption[]
{Principle of the FDPXR signal generation with the tungsten target
crystal in positive orientation (a), and negative orientation (d).
Panels (b) and (e) depict the relative energy change
$\Delta\hbar\omega/\hbar\omega_{0}$ as function of the emission
angle $\vartheta_x$ at $\vartheta_y$ = 0 for the FDPXR peak as
well as the flat Si analyzer crystal. Curves are shown for three
different variations $\psi$ = \{-0.5, 0, 0.5\} mrad of the
rotation angles with respect to the nominal direction $\Theta_0 =
3.977^\circ$, $\Theta_0^A = 2.833^\circ$. Panels (c) and (f) show
the position $x'$ of the Bragg reflex at the detector plane as
function of $\psi$. Calculations have been performed for the
experimental parameters described in the caption of Figs.
\ref{figResonance}, \ref{figsetup} and in section
\ref{sect:experimental}.}\label{figmodExpPlanePlusMinus}
\end{figure}
\begin{figure}
   \begin{center}
   \begin{tabular}{c}
   \includegraphics[height=5.5 cm]{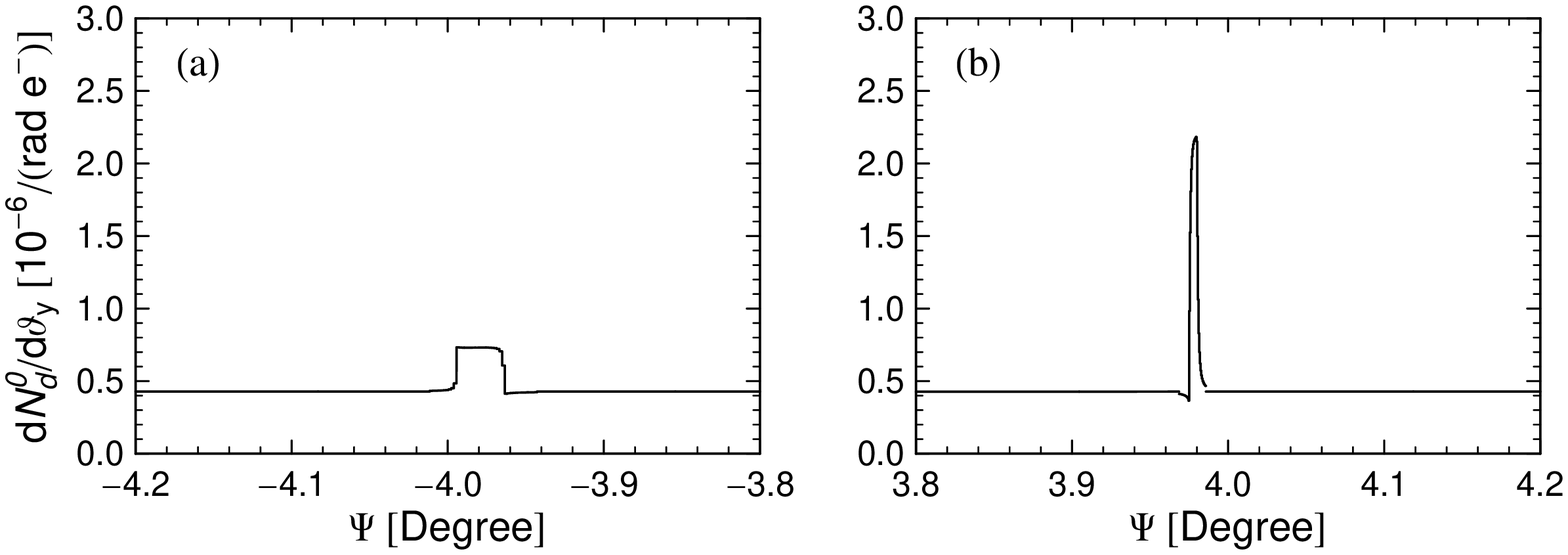}
   \end{tabular}
   \end{center}
   \caption[]
{FDPXR intensity distribution for a plane analyzer crystal,
calculated with Eq. (\ref{dN0detfinal}). Parameters of the
numerical calculation as given in captions of Figs.
\ref{figResonance}, \ref{figsetup} and in section
\ref{sect:experimental}. Fourier components of the dielectric
susceptibility for the silicon single crystal are $\chi^{ \prime
A}_{0}=-0.603 \times 10^{-6}$, $\chi^{ \prime\prime A
}_{0}=-0.533\times 10^{-9}$, $\chi^{ \prime A}_{H}=-0.316\times
10^{-6}$, $\chi^{ \prime\prime A}_{H}=-0.371\times 10^{-9}$
\cite{LucS91}.}\label{figPsiScansCalcRinfP}
\end{figure}
\subsection{Principle of the Experiment}\label{subsect:principle}
The signal generation will be explained by means of Fig.
\ref{figmodExpPlanePlusMinus}. The reciprocal lattice vector
$\textbf{H}_{0}$ of the radiator crystal as well as of the
analyzer crystal are located in the horizontal $(x,z)$ plane. The
two possibilities to place vector $\textbf{H}_{0}$ in the $(x,z)$
plane, which are shown in Figs. \ref{figmodExpPlanePlusMinus} (a)
and (d), correspond to quite different features of the FDPXR
signals. The energy of the quasi-monochromatic FDPXR line as
function of the emission angle $\vartheta_x$ is shown in panels
(b) and (e) for three different rotation angles $\psi$ of the
target crystal around the vertical $y$ axis, which are deviations
from the nominal value $\Psi = \Theta_0$. The middle line for
$\psi$ = 0 corresponds to the nominal Bragg angle
$\Psi=\pm\Theta_0$ to which a positive or a negative sign is
assigned, depending on the orientation of the $\textbf{H}_{0}$
vector. Notice, that for the positive orientation of the target
crystal the FDPXR energy increases as function of the emission
angle while for the negative orientation the energy decreases.
This feature has the consequence that the energy characteristics
of the flat analyzer crystal, which is assigned in Fig.
\ref{figmodExpPlanePlusMinus} with Si, intersects the FDPXR line
at quite different angles resulting in different intensities of
the Bragg reflex for FDPXR. The energy has been calculated with
the general relation
\begin{eqnarray}
\varepsilon_B = \Delta\hbar\omega_B/\hbar\omega_0 =
\big[(1-\frac{a}{R_0 \sin \Theta_0^{A}})~
\vartheta_x-\Delta\Theta^A \big]/\tan\Theta_0^A
 \label{bragganalyzer}
\end{eqnarray}
which holds for a cylindrically bent crystal with a bending radius
$R_0$ and takes also into account a small deviation
$\Delta\Theta^A$  from the nominal Bragg direction $\Theta_0^A$.
The quantity $a$ is the distance between target and analyzer
crystal. For a flat crystal the bending radius is
$R_0\rightarrow\infty$, and $\Delta\Theta^A = 0$. In the following
it is assumed that for $\psi$ = 0 also $\vartheta_x$ = 0 and the
reflex enters the detector slit at the position $x'$ = 0. If now
the rotation angle $\psi$ is changed also $\vartheta_x$ must be
varied in order that FDPXR and Si analyzer energies match again.
As a consequence, the reflex appears at a different position $x'$
at the detector plane. Again, the displacements are different for
the positive and the negative orientation of the W target crystal
as shown in panels (c) and (f).

In an experiment the count rate of a detector positioned behind a
vertical slit is detected as function of the rotation angle
$\psi$. The intensity is given by the double integral
\begin{eqnarray}
\frac{dN^0_{d}}{d\vartheta_{y}}\big(\psi,\vartheta_{y})=\sum_{P}\int\limits_{\vartheta_{x}}
\int\limits_{\varepsilon}\frac{d^3\overline{N}^0_{2P}}{
d\vartheta_{x}d\vartheta_{y}d\varepsilon}\big(\psi,\vartheta_{x},\vartheta_{y},\varepsilon\big)
|r_{A}^{P}(\vartheta_{x},\varepsilon)|^{2}~d\varepsilon~d\vartheta_{x}
\label{dN0det}
\end{eqnarray}
with the reflecting power ratio $|r^P_A|^2$ of the silicon
monochromator crystal. The latter was calculated from the
amplitude ratio \cite[Eq.(3.2)]{Cat89}
\begin{eqnarray}
r^P_A(u)&=&-y_P(u)+\mbox{sign}[\Re{(y_P(u))}]\sqrt{y^2_P(u)-1}, \label{rap} \\
y_P(u)&=&\frac{u+i\Im{(\chi^A_0)}}{P\chi^A_H}, \label{ypu} \\
u &=&2 \sin
\Theta^A_0\big[\widetilde{\vartheta}_x\cos\Theta^A_0+\varepsilon_B
~\sin \Theta^A_0\big]+\Re{(\chi^A_0)}. \label{u}
\end{eqnarray}
The widthes are different for the two polarization states. In
order to simplify the numerical calculations the reflecting power
ratio was approximated by a Dirac-$\delta$ function as
${|r^P_A(u)|}^2={|R^P_A|}^2 \delta(u)$ with the integrated
reflecting power $|R^P_A|^2=\int{|r^P_A(u)|^2 du}$. The numerical
values of the reflecting power ratios are ${|R^1_A(u)|}^2 =
8.154\cdot 10^{-7}$ and ${|R^{\cos2\Theta_0^A}_A(u)|}^2 =
8.114\cdot 10^{-7}$ and differ only slightly from each other for
the two polarization states at the small Bragg angle $\Theta_0^A =
2.833^{\circ}$. For a cylindrically bent crystal the angle of
incidence is given by $\widetilde{\vartheta}_x=-f_{a}\vartheta_x$
with the factor $f_{a}=1-a/(R_0 \sin \Theta_0^{A})$. With the
correlation $\varepsilon_B(\vartheta_x)=f_a \vartheta_x
/\tan\Theta_0^{A}$, which follows from Eq. (\ref{u}), Eq.
(\ref{dN0det}) can be brought into the form
\begin{eqnarray}
\frac{dN^0_{d}}{d\vartheta_{y}}\big(\psi,\vartheta_{y})=\frac{1}{2
\sin^2 \Theta^A_0} \sum_{P}|R^P_A|^2\int\limits_{\vartheta_{x}}
\frac{d^3\overline{N}^0_{2P}}{d\vartheta_{x}
d\vartheta_{y}d\varepsilon}\big(\psi,\vartheta_{x},\vartheta_{y},\varepsilon_B(\vartheta_{x})\big)
~d\vartheta_{x}. \label{dN0detfinal}
\end{eqnarray}
The integral must be taken over the angular region in
$\vartheta_{x}$ which the detector accepts. Limitations originate
from the finite analyzer crystal length and the slit aperture in
front of the detector.

The expected signals, as calculated with Eq. (\ref{dN0detfinal}),
are shown in Fig. \ref{figPsiScansCalcRinfP}. The smooth
background originates from the transition radiation contribution.

\section{EXPERIMENTAL}\label{sect:experimental}
\subsection{Experimental Setup} \label{sect:setup}
\begin{figure}[]
   \begin{center}
   \begin{tabular}{c}
   \includegraphics[width=14.0 cm]{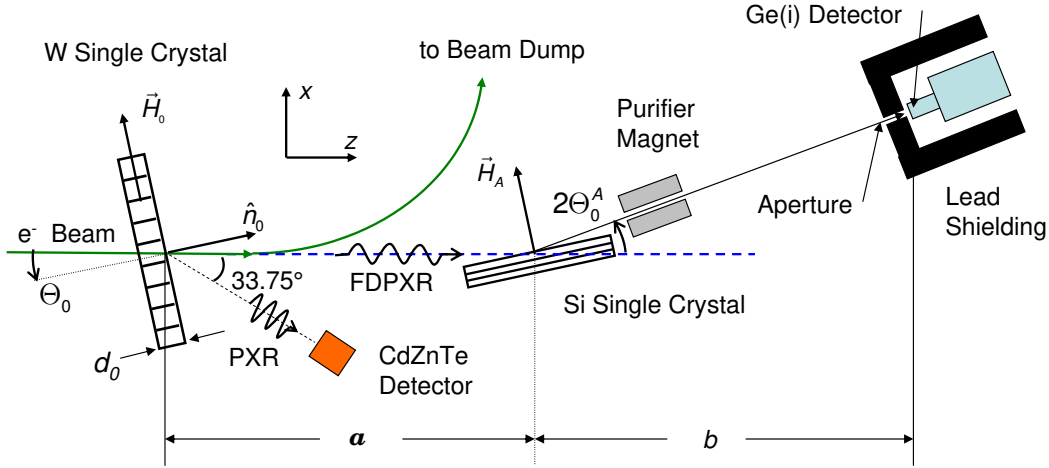}
   \end{tabular}
   \end{center}
   \caption[]
{Experimental setup at the Mainz Microtron MAMI. The flat
rectangular silicon single crystal monochromator of 150 mm length,
40 mm height and 1~mm thickness, cut with the (111) plane parallel
to the surface, was positioned in Bragg geometry at a distance of
$7.629$~m downstream the tungsten target crystal of $d_{0}$ = 410
$\mu$m thickness, and 8.5 mm diameter. Reflected radiation is
detected with a Ge(i) detector with 6.18 mm diameter and 5 mm
thickness positioned in a distance $b$ = 8.436 m from the analyzer
crystal. The slit aperture had a width of 3.5 mm in horizontal
direction. The lead shielding of the Ge(i) detector had a
thickness of 50 mm. The purifier magnet with a gap width of 60 mm
and a length of 370 mm deflects with its magnetic field of 0.12
Tesla shower electrons and positrons created in the silicon single
crystal in vertical direction. Ordinary PXR was detected with a 3
mm $\times$ 3mm $\times$ 2mm CdZnTe detector which was positioned
in a distance of 350 mm from the target at an angle of
33.75$^{\circ}$ in the horizontal $(xy)$ plane.}\label{figsetup}
\end{figure}

The basic idea of our experiment was already explained in
subsection \ref{subsect:principle}. The details of the setup will
be described in the following with the help of Fig.
\ref{figsetup}. The
target tungsten single crystal, cut with the (111) plane parallel
to the crystal surface, must be positioned in such a manner that
the reciprocal lattice vector $\textbf{H}_{0}$ of the
$(10\bar{1})$ crystal plane is located in the plane of drawing. At
a Bragg angle $\Theta_{0} = 3.977 ^{\circ}$ the photon energy
amounts to $\hbar\omega_{0} = 40.0$ ~keV for both, PXR emitted at
twice the Bragg angle and FDPXR emitted close to the electron beam
direction. The required energy resolution in the order of $1\cdot
10^{-4}$ and at the same time a good angular resolution is
achieved by a flat silicon single crystal in Bragg geometry. The
monochromator crystal was cut with the (111) plane parallel to the
crystal surface and was used to analyze the FDPXR with its (111)
reflection. At a Bragg angle $\Theta_{0}^{A} = 2.833^{\circ}$ the
analyzing energy amounted also to $\hbar\omega_{0} = 40.0$ ~keV.
The angular width of this reflection for a photon line with an
assumed $\delta$-function-like shape is about 7 $\mu$rad, the
energy width for an infinitesimal small accepted angular band
about 1.4 $10^{-4}$. These numbers match well with the expected
multiple scattering broadened FDPXR peak, see Figure
\ref{figResonance}. The observation angle $\vartheta_{x}\simeq 0$
is selected by a slit of 3.5 mm width in front of the photon
detector. At target-monochromator and monochromator-detector
distances of 7629 mm and 8436 mm, the angular interval accepted by
such a slit aperture is $\Delta\Theta_{x}$ = 0.22 mrad which is
somewhat larger as the calculated FDPXR width of 14 $\mu$rad.
However, it should be mentioned that the energy-angle correlations
of both, FDPXR and analyzer crystal, defines also the accepted
angular band. The FDPXR resonance is sought by a variation of the
rotation angle $\psi$ of the tungsten crystal around the vertical
$y$ axis. As photon detector a Ge(i) detector with a resolution of
630 eV at 40 keV was used.

\subsection{Measurements and Results}\label{sect:results}
The experiments were performed at an electron beam energy of
855~MeV delivered by the Mainz Microtron MAMI. The beam current
was about 2~nA. The beam spot size amounted to 48 $\mu$m (rms)
horizontally and 55 $\mu$m (rms) vertically. At a beam emittance
of $7 \cdot 10^{-9}$ m$\cdot$rad and $0.5 \cdot 10^{-9}$
m$\cdot$rad the angular divergence of the electron beam was 146
$\mu$rad (rms) and 9 $\mu$rad (rms) in horizontal and vertical
direction, respectively. These numbers are small in comparison to
the opening angle of the TR cone which is in the order of
1/$\gamma$ = 0.6 mrad. The [111] direction of the tungsten target
crystal was aligned into the beam direction by means of a
goniometric stage with angular resolution of $(2/1,000)^{\circ}$.
A signal which is sensitive to channeling was derived from an
ionization chamber located 400 mm behind the analyzer crystal.
After the [111] direction of the crystal was found, the reciprocal
lattice vector $\textbf{H}$ for one of the three $(10\bar{1})$
crystal planes had to be placed into the horizontal $(xz)$ plane.
This was achieved by observation of the ordinary PXR with the aid
of CdZnTe detector, see Fig. \ref{figsetup}. The target crystal
was rotated around the vertical $y$ axis by an angle
$\Psi=(33.75/2)^{\circ}$ in order to fulfill the Bragg condition.
The PXR photon energy at this angle is 9.6 keV. At a rotation
around the [111] crystal axis from 0-360$^{\circ}$ six PXR
reflexes of equal intensity are expected which are separated by
60$^{\circ}$ from each other. However, we observed rather
irregular intensities of these reflexes. The reason was found in
the fact that the surface normal of the crystal
$\hat{\textbf{n}_0}$ and the reciprocal lattice vector
$\hat{\textbf{H}_0}$ for the [111] direction obviously do not
coincide. Since the goniometric stage rotates the crystal around
the $\hat{\textbf{n}_0}$ axis, the Bragg angle varies at the
rotation, i.e. the angle between beam direction and the various
$\hat{\textbf{H}_0}$ vectors of the $(10\bar{1})$ family. The
experimental intensities could be explained with the assumption
that rotation axis of the goniometer and [111] direction of the
crystal make an angle of 1.69$^{\circ}$. At a rotation of the
crystal the Bragg angles are 16.80$^{\circ}$, 15.12$^{\circ}$,
14.11$^{\circ}$, 14.78$^{\circ}$, 16.46$^{\circ}$, and
17.47$^{\circ}$.  Only two values are close to the nominal Bragg
angle of 16.88$^{\circ}$ in accord with the experimental
observation that only two strong lines and four rather weak ones
were observed. The settings of the goniometer found for the
strongest reflection was chosen.

A very simple check that the right $(10\overline{1})$ lattice
planes were found can be performed by a small displacement of the
detector in $x'$ direction, re-adjustment of the analyzer Bragg
angle by the corresponding angle $\Delta\Theta^A$ and a
measurement of the angle $\psi$ at which the FDPXR peak reappears
again. From Eqns. (\ref{e0}) and (\ref{bragganalyzer}) it follows
that the ratio of these angles is $\psi/\Delta\Theta^A|_{calc} =
\tan\Theta_0^A/\tan\Theta_0 = \pm 1.405$ with the plus sign for
the positive and the minus sign for the negative orientation.
Experimentally we found $\psi/\Delta\Theta^A|_{exp} = 1.34\pm
0.14$ mrad/mrad and $-1.43\pm 0.14$ mrad/mrad, respectively.

\begin{figure}[!]
   \begin{center}
   \begin{tabular}{c}
   \hspace{2.0 mm} \includegraphics[width=14.57 cm]{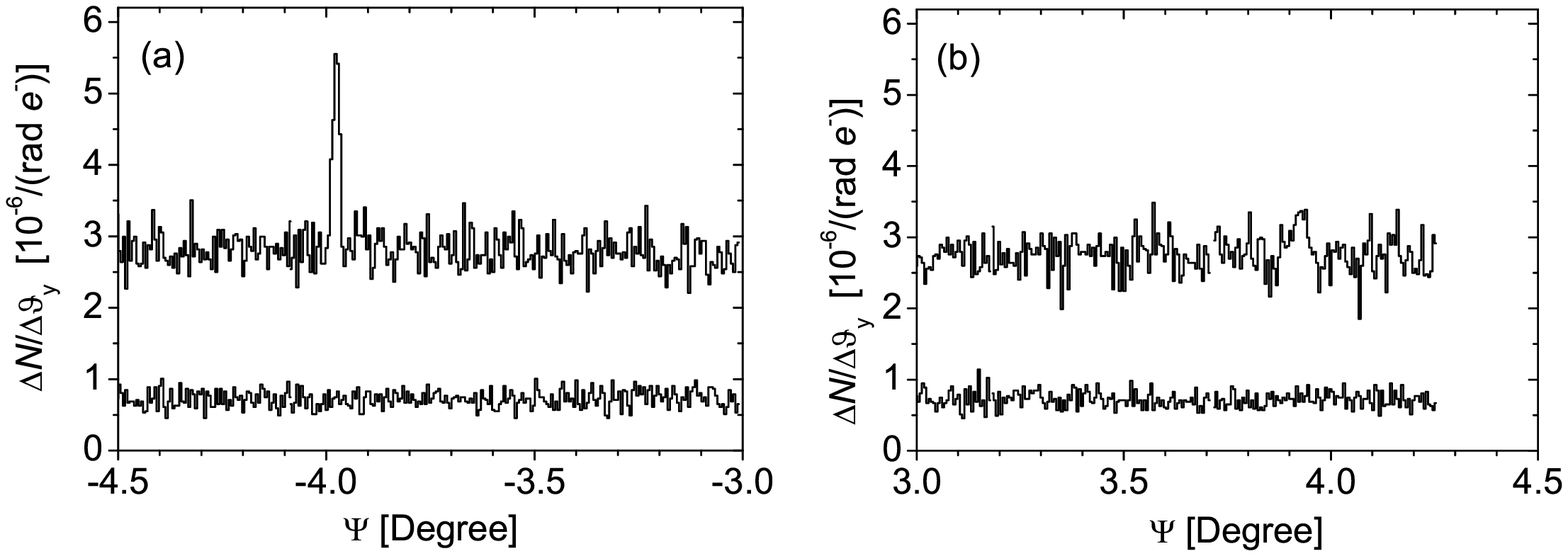}\\
   \includegraphics[width=15.0 cm]{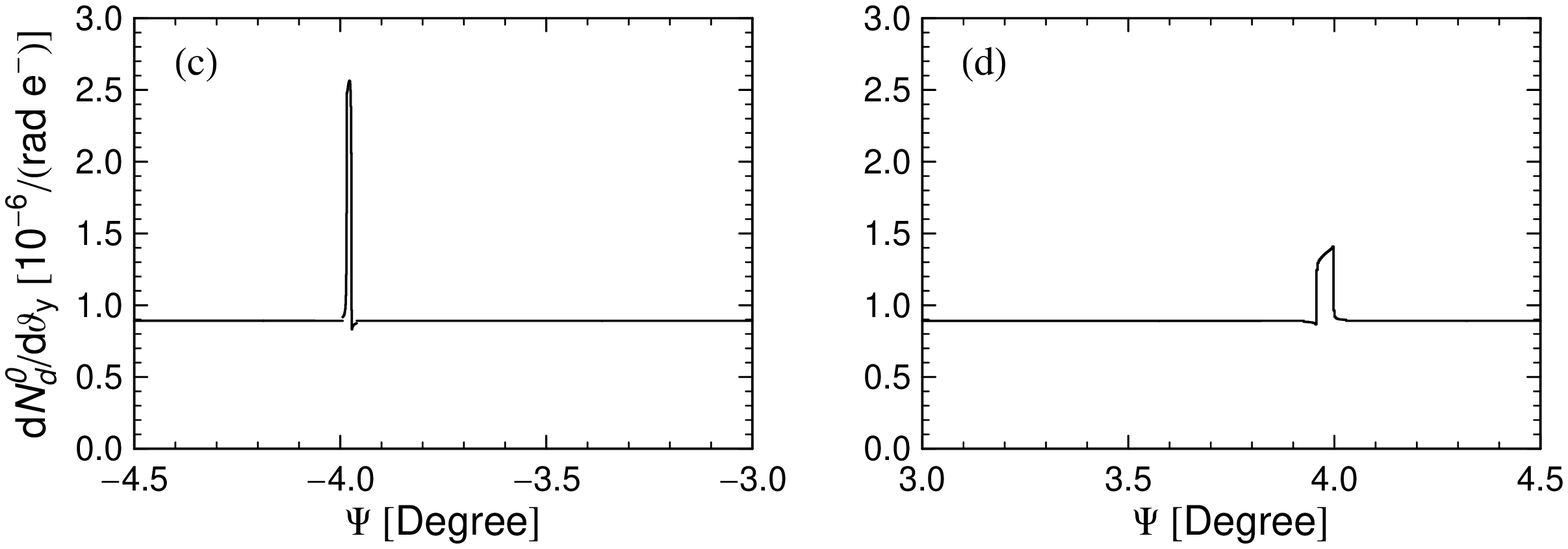}
   \end{tabular}
   \end{center}
   \caption[]
{Measured and calculated scans of the tungsten crystal around the
vertical $y$ axis. Panels (a) and (b) are experimental results.
The fraction of the background originating from bremsstrahlung is
indicated (lower curves), but not subtracted. Panels (c) and (d)
show calculations assuming a concave cylindrical shape of the
analyzing crystal with a bending radius $R_0$ = 110
m.}\label{figPsiScansExpCalcR110}
\end{figure}

For the search of the FDPXR peaks angular scans around the
vertical $y$ axis were carried out in the angular interval
$-4.5^\circ<\Psi<-3.0^\circ$ and $3.0^\circ<\Psi<4.5^\circ$. The
scans are shown in Figure \ref{figPsiScansExpCalcR110} (a) and
(b).
\begin{figure}[!]
   \begin{center}
   \begin{tabular}{c}
   \includegraphics[width=13.0 cm]{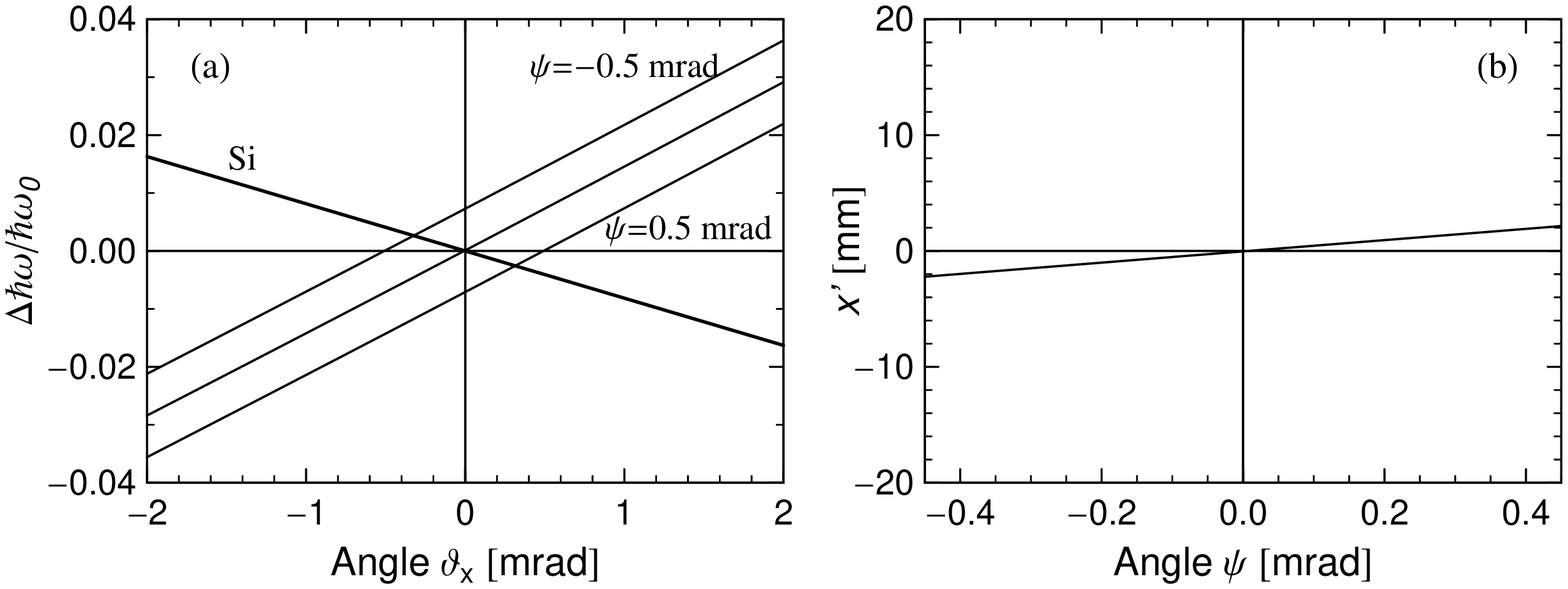}\\
   \includegraphics[width=13.0 cm]{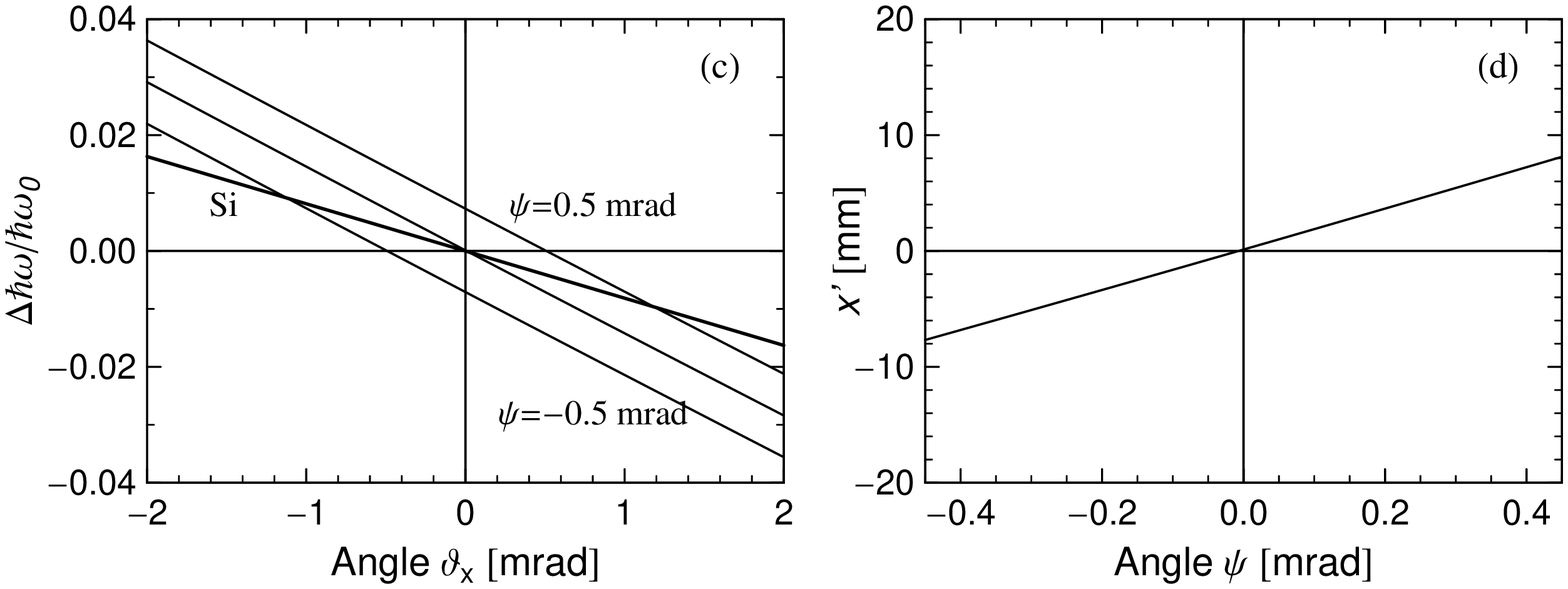}
   \end{tabular}
   \end{center}
   \caption[]
{FDPXR signal generation with the tungsten target crystal in
positive and negative orientation for a concave cylindrical shape
of the analyzing crystal with a bending radius $R_0$ = 110 m.
Panels (a) and (b) depict calculations for positive, (c) and (d)
for negative orientation of the tungsten target crystal. Shown are
in panels (a) and (c) the relative energy changes
$\Delta\hbar\omega/\hbar\omega_{0}$ as function of the emission
angle $\vartheta_x$ at $\vartheta_y$ = 0 for the FDPXR peak as
well as the flat Si-analyzer crystal assigned with "Si". Panels
(b) and (d) show the position $x'$ of the Bragg reflex at the
detector plane. For further explanations see caption of Fig.
\ref{figmodExpPlanePlusMinus}.}\label{figmodR110PlusMinus}
\end{figure}

\section{Discussion}\label{sect:discussion}
The striking result of the experiment described in section
\ref{sect:experimental} is the fact that for the negative
orientation a clear peak with a peak-to-background ratio of 1:1
was observed, while for the positive orientation the peak is
rather weak. This experimental finding is just the opposite of
what was expected, see Fig. \ref{figPsiScansCalcRinfP}. A possible
experimental mistake in the assignment of positive and negative
orientation of the tungsten crystal was carefully checked and
could safely be excluded as an explanation. Also the signs in the
FDPXR formulas underlying the calculations were checked and a
mistake could not be found. The only reasonable explanation for
the reversed intensities was found in a residual bending of the
"plane" silicon analyzer crystal. Assuming a cylindrical shape of
the analyzing crystal with a bending radius $R_0$, the energy
characteristics of the analyzer crystal changes its slope as
function of $R_0$. As shown in Fig. \ref{figmodR110PlusMinus} a
reversed situation in comparison to Figs.
\ref{figmodExpPlanePlusMinus} can be achieved with a concave
cylindrical shape of the analyzing crystal and a bending radius
$R_0$ = 110 m. The corresponding calculated scans are shown in
Fig. \ref{figPsiScansExpCalcR110} (c) and (d). The line widthes
are narrower and more intense. However, this fact must not be
overrated since a number of line broadening effects have been
disregarded as, e.g., the beam spot size, tungsten- and
analyzer-crystal irregularities, etc. In addition, also the width
of the aperture in front of the Ge(i) detector has an uncertainty
which originates from misalignments of a 50 mm long lead aperture
with a bore of 8 mm diameter by an angle of about 4$^\circ$.

This possible explanation prompted us to carry out experiments
with a silicon single crystal with a length of 100 mm, a height of
50 mm and a thickness of 10 mm which was assumed to be really
plane. The experimental setup was a little modified. The distances
between W target-analyzer-detector were symmetrized and selected
as $a = b = 7.629$ m. A vertical slit aperture of 0.2 mm width was
positioned just in front of the Si analyzer crystal. This aperture
could horizontally be moved during the course of the experiment
enabling us to investigate a possible local bending radius change
over the crystal. In addition, the slit aperture in front of the
Ge(i) detector was reduced to a width of 1 mm and the detector
assembly was made moveable in $x'$ direction. With this setup we
found a quite strange behavior which we interpreted as rapidly
changing wavy structures of the (111) lattice planes across the
length of the Si crystal resembling somehow a mosaic structure. As
a result, further experiments with this "plane" crystal were
abandoned.

\begin{figure}[!]
   \begin{center}
   \begin{tabular}{c}
   \includegraphics[width=6.693 cm,bbllx=31,bblly=30,bburx=577,bbury=755]{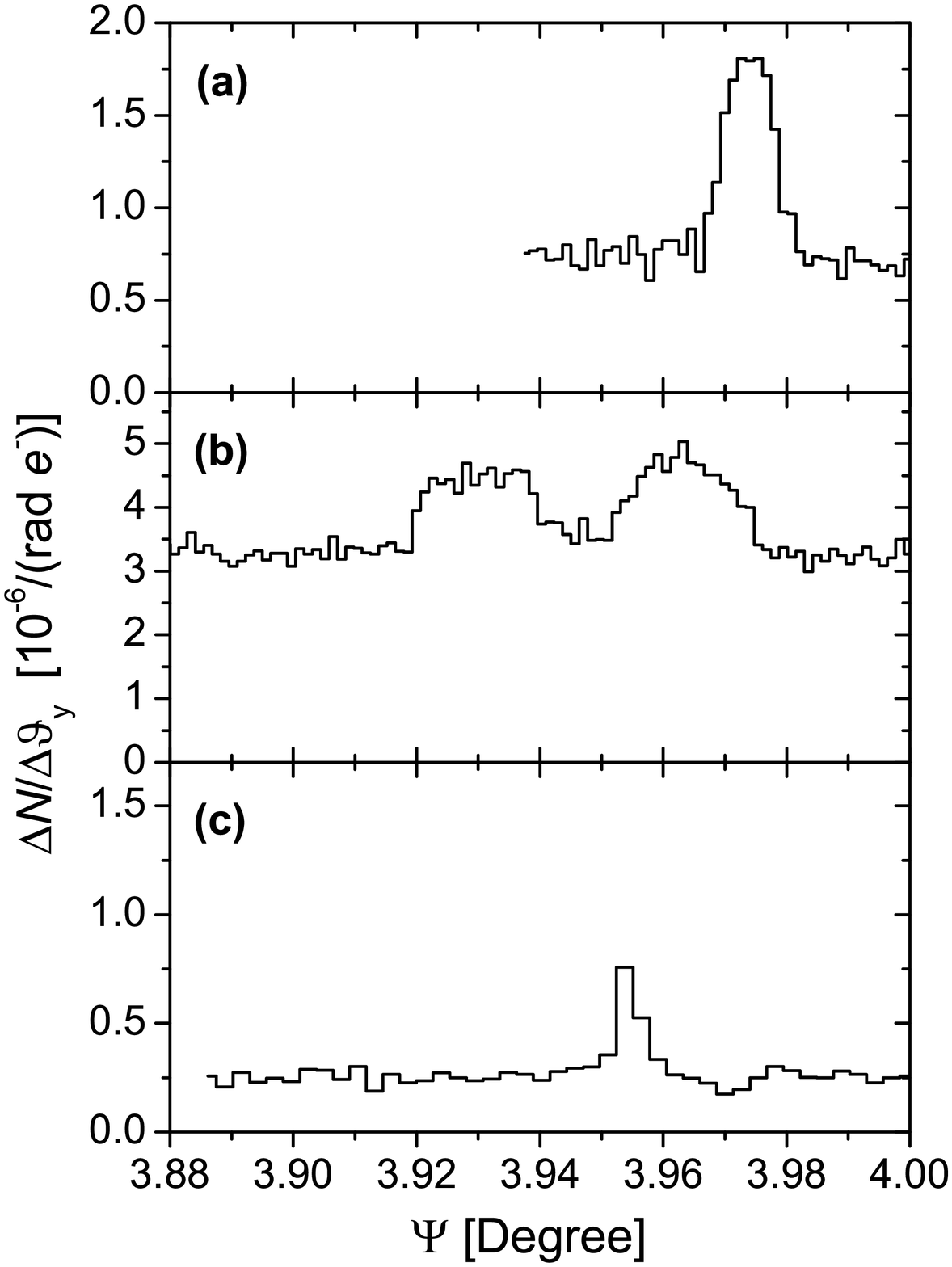}
   \includegraphics[width=7.0 cm]{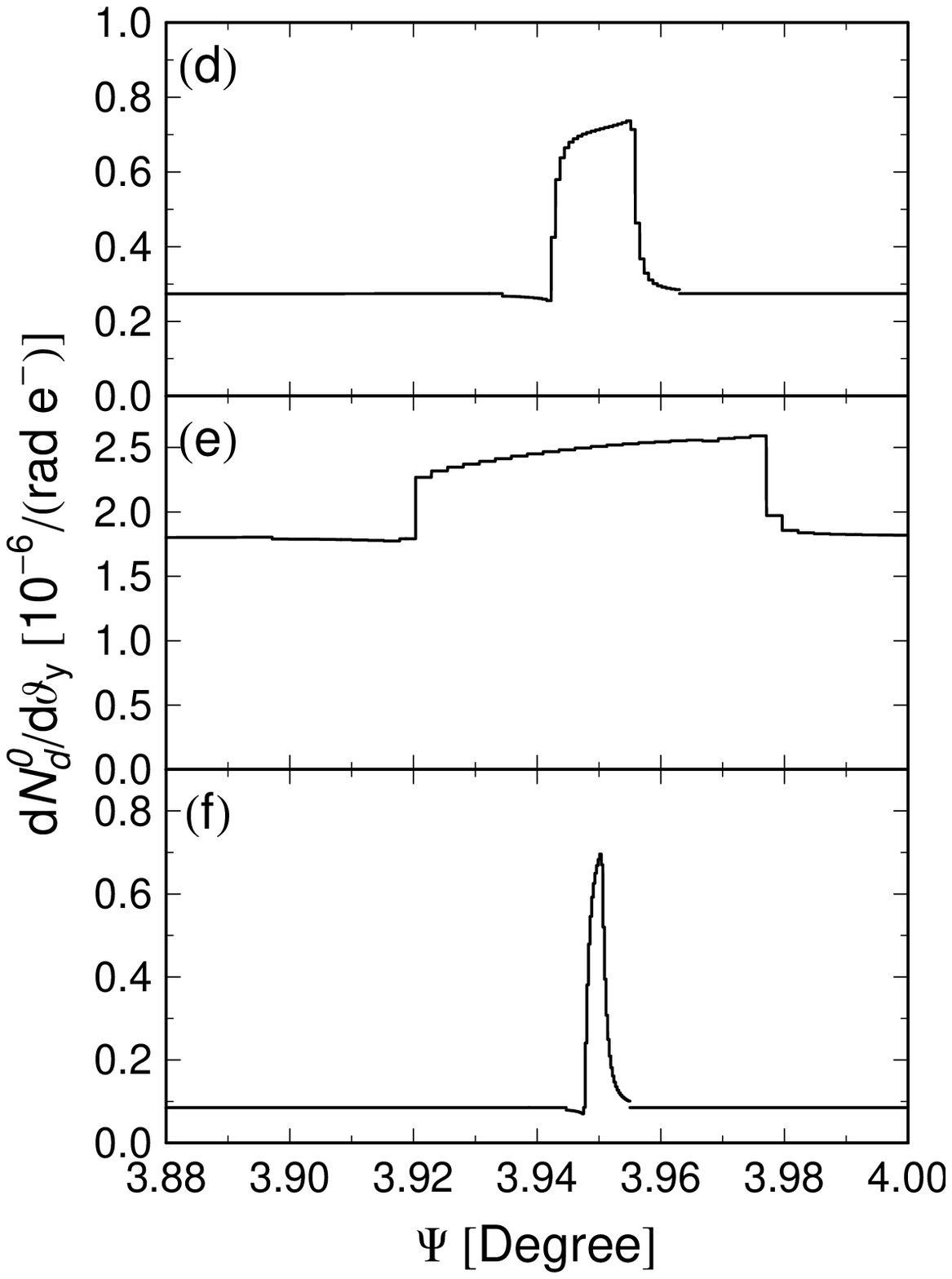}
   \end{tabular}
   \end{center}
   \caption[]
{FDPXR characteristics as function of the bending radius of the
Si-monochromator crystal. Experimental $\Psi$ scans are shown in
panels (a), (b), (c), bremsstrahlung background was subtracted,
calculated ones in (d), (e), (f). Target-monochromator-detector
distances were $a = b = 7.629$ m, for definitions see Fig.
\ref{figsetup}. Panel (a) shows a measurement without any tension
exerted on the Si-monochromator crystal, panel (b) for a bending
that the target radiation was focused as good as possible onto the
1 mm wide slit aperture in front of the Ge(i) detector, and (c)
with still increased tension. Panel (d) depicts a calculations
with a bending of the Si- monochromator crystal of $R_0$ = 105 m,
panel (e) with $R_0$ = 150 m, panel (f) with $R_0$ = -300 m, i.e
the crystal has a convex shape.} \label{figPsiScansExpCalc0706}
\end{figure}

Finally, we performed experiments with a third crystal again with
a length of 150 mm, a height 40 mm and a thickness of 1 mm. The
crystal was clamped on one of the shorter edges and could be
distorted with a moveable pin which touched the crystal at the
opposite side. This way the bending radius could be changed,
though not homogeneously over the whole crystal since the crystal
had not a triangular shape. All other experimental conditions were
chosen as described above. In a first step of the experiment a
scan was taken of the relaxed crystal. The result is shown in Fig.
\ref{figPsiScansExpCalc0706} (a). The corresponding calculation of
Fig. \ref{figPsiScansExpCalc0706} (d) indicates that also this
crystal has a residual concave bending with a radius $R_0 = 105$
m. In a second step of the experiment the focus was searched for
by changing systematically the bending radius and observing the TR
intensity as function of the detector position $x'$. After the
maximum was found a $\Psi$ scan of the W-target crystal in
positive orientation was carried out. The result is shown in Fig.
\ref{figPsiScansExpCalc0706} (b). The width of the FDPXR structure
of 0.96 mrad matches well with the projection of the Si-analyzer
crystal length $l_c$ = 150 mm on the $x$ axis. The corresponding
accepted angle is $l_c \sin\Theta_0^A$/a = 0.97 mrad. (Notice,
that in Eq. (\ref{e0}) which describes the energy of the FDPXR
peak $\vartheta_x$ and -$\psi$ appear symmetrically.) Most
remarkable is the double peak structure. Two additional scans of
the aperture in front of the Si crystal for fixed $\Psi$ values,
chosen in the maxima of both peaks, revealed that they are
correlated to two distinct $\vartheta_x$ regions, each of about
0.13 mrad width and separated by 0.39 mrad. This finding suggests
that the 1 mm aperture in front of the detector was not wide
enough to accept all rays reflected by the Si analyzer crystal.
Obviously rays from certain $\vartheta_x$ regions are excluded
from detection by the aperture. This conjecture was corroborated
with calculations on the basis of our FDPXR model.

In the model calculations we adjusted the reciprocal lattice
vector and the surface normal to $\hat{\textbf{H}_0}$= (0.997588,
-0.00872654, -0.0688609) and $\hat{\textbf{n}_0}$= (0.0685897,
-0.0400394, 0.996841), respectively, in order to take into account
the experimental shift of the FDPXR peak from 3.977$^\circ$ to
$\Psi = 3.95^\circ$. The results of the calculations are shown in
Fig. \ref{figPsiScansExpCalc0706} (d), (e), and (f). In panel (e),
for which a bending radius $R_0$ = 150 m was assumed, all of the
radiation impinging on the Si crystal is focused into the 1 mm
slit aperture and the FDPXR structure turns out to be rather flat.
However, if an additional deformation is superimposed on the
circular bending of the crystal, e.g. a sinusoidal shape with a
period of the crystal length of 150 mm and an amplitude of only 1
$\mu$m, a dip in the middle can be produced which resembles the
dip in the measurement shown in panel (b).

A peculiar situation occurs if the bending radius of the Si
analyzer crystal is chosen such that the slope of FDPXR and the
analyzer energy characteristics are just the same, see Figs.
\ref{figmodExpPlanePlusMinus} and \ref{figmodR110PlusMinus} for
illustration. This happens for bending radii $R_0$ = 535.52 m and
90.17 m for the positive and negative orientation of the W single
crystal, respectively. For these magic radii the complete FDPXR
spectrum will be focused on a single spot which is located for our
symmetrical geometry according to the image equation
$1/a+1/z_i=1/f$, with the focal length $f = (R_0/2)
\sin\Theta_0^A$, at distances of $z_1$ = -18.01 m and $z_2$ = 3.15
m from the Si analyzer crystal, i.e. the former is a virtual the
latter a real focus. If the above condition is fulfilled the
intensity originating from normal TR will be enhanced by the FDPXR
contribution within an angular tuning range of the target crystal
in the order of $\Delta\Psi$ = 10 $\mu$rad, see Fig.
\ref{figResonance}, which now must be convoluted with the
reflecting power ratio of the analyzer crystal $|r^P_A|^2$, Eq.
(\ref{rap}). Since the analyzer crystal has never the ideal
structure as mathematically assumed, on this basis a method could
be developed to diagnose the surface bending radii of large area
single crystals. In order to get familiar with the properties of
the FDPXR radiation which is diffracted by the Si analyzer crystal
we decreased the bending radius. An example is shown in Fig.
\ref{figPsiScansExpCalc0706} (c). As shown in panel (f) of this
Figure \ref{figPsiScansExpCalc0706} the experimental result is in
accord with the assumption of a bending radius of minus 300 m,
i.e. the crystal was bent from the concave into a convex shape.
Alternatively, it could also be that a piece of the crystal acted
as monochromator which had intrinsically a concave shape.
Unfortunately, the beam-time allocated for the experiment was
exhausted after this scan was taken and a search for the magic
bending radii could not be carried out anymore.

\section{Conclusions and Outlook}\label{sect:conclusion}
Clear forward diffracted PXR structures have been observed for a
410 $\mu$m thick tungsten single crystal which was irradiated with
the 855 MeV electron beam of the Mainz Microtron MAMI. For design
and interpretation of the experiments the results of a dynamical
formalism developed in Mainz, which predicts FDPXR from a
semi-infinite crystal quite accurately and from crystal slabs
approximately, were of crucial importance. From a comparison of
the model predictions and the experimental results it can safely
be concluded that PXR production is a dynamical rather than a
kinematical process. The experimental procedure may be developed
further to diagnose local crystal bending radii as large as 1 km
of large area crystals.

An interesting question is whether TR from the entrance interface
of the crystal may penetrate the 410 $\mu$m thick crystal. The
absorption length at a photon energy of 40 keV for amorphous
tungsten amounts to 53.4 $\mu$m \cite{LucS91} and only a fraction
of $4.6\cdot10^{-4}$ of the radiation should transmit the crystal.
However, the transmission may be much larger in a single crystal
close to a resonance where the absorption length may be
anomalously large \cite{ImaN01}. In a next step, the (333)
reflection of tungsten will be analyzed, which has been measured
simultaneously with the (333) reflection of the Si-analyzer
crystal. At a photon energy of 120 keV the absorption length
amounts to 204.3 $\mu$m \cite{LucS91} and a large fraction of 13.4
\% of TR from the entrance interface transmits the crystal which
superimposes with TR and FDPXR produced at the exit interface.

\acknowledgments

We gratefully acknowledge support of M. Tabrizi and T. Weber
during the course of the experiment. View graphs in the
Mathematica 5.1 environment were generated with the LevelScheme
figure preparation system of Ref. \cite{Cap05} version 3.21
(October 23, 2005).

This work has been supported by Deutsche Forschungsgemeinschaft
DFG under contract BA 1336/1-4.


\begin{thebibliography}{1}  

\bibitem {GinF45} V.L. Ginsburg, I.M. Franck, {\em J. Phys. (Moscow)} {\bf IX}, p. 353, 1945.

\bibitem {Gar58} G.M. Garibian, ``Contribution to the theory of
transition radiation'', {\em Zh. Exper. Teor. Fiz.} {\bf 33}
1403-1410, 1957; {\em Sov. Phys. JETP} {\bf 6}, p. 1079-1085,
1958.

\bibitem {Gar61} G.M. Garibian, ``Radiation of a particle moving
across the interface of two media with account of multiple
scattering'', {\em Zh. Exper. Teor. Fiz.} {\bf 39} 332-336, 1960;
{\em Sov. Phys. JETP} {\bf 12}, p. 237-239, 1961.

\bibitem {Cat89} A. Caticha, ``Transition-diffracted radiation and the Cerenkov emission of x rays'',
{\em Phys. Rev.} {\bf A 40}, p. 4322-4329, 1989.

\bibitem {Ter72} M. Ter-Mikaelian, High-Energy Electromagnetic Processes
in Condensed Media, Wiley-Interscience, New York, London, Sydney,
Toronto, 1972.

\bibitem {Bar71} V.G. Baryshevsky, {\em Doklady Akad. Nauk BSSR} {\bf
15} p. 306, 1971.

\bibitem {BarF71} V.G. Baryshevsky and I.D. Feranchuk,``Transition
radiation of $\gamma$ rays in a crystal'', {\em Zh. Exper. Teor.
Fiz.} {\bf 61} 944-948, 1971;  ({\em Sov. Phys. JETP} {\bf 34}, p.
502-504, 1972; addendum, {\em ibid} {\bf 64}, p. 760, 1973).

\bibitem {GarY71} G.M. Garibian and C.Yang,
``Quantum microscopic theory of radiation by charged particle
moving uniformly in a crystal'', {\em Zh. Eksp. Teor. Fiz.} {\bf
61}, p. 930-943, 1971, ({\em Sov. Phys. JETP} {\bf 34}, p.495-501,
1972).

\bibitem {GarY72} G.M. Garibian and C.Yang, ``Lateral spots of X-ray
transition radiation in crystals and their effect on the central
spot'', {\em Zh. Eksp. Teor. Fiz.} {\bf 63}, p. 1198-1211, 1972,
({\em Sov. Phys. JETP} {\bf 36}, p. 631-637, 1973).

\bibitem {GarY86} G.M. Garibian and C. Yang,``Quasi-Cherenkov radiation in
crystals'', {\em Nucl. Inst. Meth.} {\bf A 248}, p. 29-30, 1986.

\bibitem {Nit91} H. Nitta, ``Kinematical Theory of parametric X-ray
radiation'', {\em Phys. Lett.} {\bf A 158}, p. 270-274, 1991.

\bibitem {Nit92} H. Nitta, ``Theory of coherent X-ray radiation by
relativistic particles in a single crystal'', {\em Phys. Rev.}
{\bf B 45}, p. 7621-7626, 1992.

\bibitem {FreG95} J. Freudenberger, V.B. Gavrikov, M. Galemann, H.
Genz, L. Groening, V.L. Morokhovskii, V.V. Morokhovskii, U.
Nething, A. Richter, J.P.F. Sellschop, N.F. Shulga, ``Parametric
x-ray radiation observed in diamond at low electron energies'',
{\em Phys. Rev. Lett.} {\bf 74}, p. 2487-2490, 1995.

\bibitem {Cat92} A. Caticha, ``Quantum theory of the dynamical Cerenkov emission of
x rays'', {\em Phys. Rev.} {\bf B 45}, p. 9541-9550, 1992.

\bibitem {ArtR01} X. Artru, P. Rullhusen,``Parametric X-rays and
diffracted transition radiation in perfect and mosaic crystals'',
{\em Nucl. Instr. Meth. in Phys. Res.} {\bf B 145}, p. 1-7, 1998;
addendum, {\em ibid} {\bf B 173}, p. 16, 2001.

\bibitem {BreH97} K.-H. Brenzinger, C. Herberg, B. Limburg, H.
Backe, S. Dambach, H. Euteneuer, F. Hagenbuck, H. Hartmann, K.
Johann, K.H. Kaiser, O. Kettig, G. Knies, G. Kube, W. Lauth, H.
Sch{\"o}pe, Th. Walcher, ``Investigation of the production mechanism
if parametric X-ray radiation'', {\em Z. Phys.} {\bf A 358}, p.
107-114, 1997.

\bibitem{FreG97} J. Freudenberger, H. Genz,
V.V. Morokhovskii, A. Richter, V.L. Morokhovskii, U. Nething, R.
Zahn, J.P.F. Sellschop, ``Lineshape, linewidth and spectral
density of parametric x-radiation at low electron energy in
diamond'', {\em Appl. Phys. Lett.} {\bf 70}, p. 267-269, 1997.

\bibitem{BreL97} K.-H. Brenzinger, B. Limburg, H. Backe, S. Dambach,
H. Euteneuer, F. Hagenbuck, C. Herberg, K.H. Kaiser, O. Kettig, G.
Kube, W. Lauth, H. Sch{\"o}pe, Th. Walcher, ``How narrow is the
linewidth of parametric X-ray radiation?'', {\em Phys. Rev. Lett.}
{\bf 79}, p. 2462-2465, 1997.

\bibitem{MorS97}
V.V. Morokhovskii, K.H. Schmitt, G. Buschhorn, J. Freudenberger,
H. Genz, R. Kotthaus, A. Richter, M. Rzepka, P.M. Weinmann,
``Polarization of parametric X radiation'', {\em Phys. Rev. Lett.}
{\bf 79}, p. 4389-4392, 1997.

\bibitem{FreG00} J. Freudenberger, H. Genz,
V.V. Morokhovskii, A. Richter, J.P.F. Sellschop, ``Parametric X
rays observed under Bragg condition: Boost of intensity by a
factor of two'', {\em Phys. Rev. Lett.} {\bf 84}, p. 270-273,
2000.

\bibitem {Nit00} H. Nitta,``Dynamical effect on parametric X-ray radiation'',
{\em Journ. Phys. Soc. Japan} {\bf 69}, p. 3462-3465, 2000.

\bibitem {Bar97} V.G. Baryshevsky, ``Parametric X-ray radiation at a small angle near
the velocity direction of the relativistic particle'', {\em Nucl.
Instr. Meth. in Phys. Res.} {\bf B 122}, p. 13-18, 1997.

\bibitem {KubN03} A. Kubankin, N. Nasonov, V. Sergienko, I.
Vnukov, ``An investigation of the parametric X-rays along the
velocity of emitting particle'', {\em Nucl. Instr. Meth. in Phys.
Res.}, {\bf B 201}, p. 97-113, 2003.

\bibitem {NasN03} N. Nasonov, N. Noskov, ``On the parametric X-rays along
an emitting particle velocity'', {\em Nucl. Instr. Meth. in Phys.
Res.} {\bf B201}, p. 67-77, 2003.

\bibitem {YuaA85} L.C.L. Yuan, P.W. Alley, A. Bamberger, G.F. Dell, H.
Uto,``A search for dynamic radiation from crystals'', {\em Nucl.
Instr. Meth. in Phys. Res.} {\bf A 234}, p. 426-429, 1985.

\bibitem {KalN01} B.N. Kalinin, G.A. Naumenko, D.V. Padalko,
A.P. Potylitsin, I.E. Vnukov,``Experimental search of parametric
X-ray radiation in a silicon crystal at a small angle near the
velocity direction of relativistic electrons'', {\em Nucl. Instr.
Meth. in Phys. Res.} {\bf B 173}, p. 253-261, 2001.

\bibitem {BacB94} H. Backe, V.G. Baryshevsky, Th. Doerk, Th. Kerschner,
H. Koch, G. Kube, W. Lauth, H. Matth{\"a}y, M. Sch{\"u}ttrumpf, A.
Wilms, M. Zemter, {\it unpublished}.

\bibitem {KerT98} Th. Kerschner, Entwicklung und Aufbau eines
pn-CCD-Systems zum Einzelphotonnachweis im Bereich weicher
R{\"o}ntgenstrahlung, Dissertation am Institut f{\"u}r
Experimentalphysik I der Ruhr-Universit{\"a}t Bochum, Bochum,
1998.

\bibitem {BacA03} H. Backe, C.Ay, N. Clawiter, Th. Doerk, M. El-Ghazaly,
K.-H. Kayser, O. Kettig, G. Kube, F. Hagenbuck, W. Lauth, A.
Rueda, A. Scharafutdinov, D. Schroff, T. Weber in: W Greiner,
A.Solov'yov and S. Misicu (Eds.),``Diffracted transition radiation
and parametric X radiation from silicon single crystal slabs'',
{\em Proc. Symp. Channeling - Bent Crystals - Radiation Processes,
Frankfurt (Germany) 2003, EP Systema, Debrecen}, p. 41-58, 2003.

\bibitem{AleB04}
N. Aleinik, A.N. Baldin, E.A. Bogomazova, I.E. Vnukov, B.N.
Kalinin, A.S. Kubankin, N.N. Nasonov,G.A. Naumenko, A.P.
Potylitsin, A.F. Scharafutdinov, ``Experimental Observation of
Parametric X-Ray Radiation Directed Along the Propagation
Direction Velocity of Relativistic Electrons in a Tungsten
Crystal,''{\em JETP Lett.} {\bf 80}, pp.~393-397, 2004 (Zh. Exper.
Teor. Fiz. {\bf 80}, pp. 447-451, 2004).

\bibitem{BacR05}
H. Backe, A. Rueda, W. Lauth, N. Clawiter, M. El-Ghazaly, P. Kunz,
T. Weber, ``Forward Diffracted Parametric X Radiation from a
Silicon Single Crystal,'' {\em Nucl. Instr. Meth. in Phys. Res.}
{\bf B 234}, p.~138-147, 2005.

\bibitem {BacR03} H. Backe, W. Lauth, A. Rueda et al., to be
published.

\bibitem {LucS91}
O.M. Lugoskaya, and S.A. Stephanov, ``Calculation of the
polarizabilities of crystals for diffraction of x-rays of the
continuous spectrum at wavelengths of 0.1-10 $\AA$'', {\em Sov.
Phys. Crystallogr.} {\bf 36}, p. 478-481, 1991, and
http://sergey.gmca.aps.anl.gov/cgi/X0h.html

\bibitem {LynD91} G.R. Lynch and O.I. Dahl, ``Approximations to multiple Coulomb scattering'',
{\em Nucl. Instr. Meth. in Phys. Res.} {\bf B 58}, p. 6-10, 1991

\bibitem {Cap05} M.A. Capiro,``LevelScheme: A level scheme drawing and
scientific figure preparation system for Matematica'', {\em
Comput. Phys. Commun.} {\bf 171}, p. 107-118, 2005.
http://wnsl.physics.yale.edu/levelscheme/

\bibitem{ImaN01} N. Imanishi, N. Nasonov, K. Yajima, ``Dynamical diffraction
effects in the transition radiation of a relativistic electron
crossing a thin crystal'', {\em Nucl. Instr. Meth. in Phys. Res.}
{\bf B 173}, p. 227-237, 2001.

\end{thebibliography}
\end{document}